\documentclass[aps,prb,floatfix,preprint]{revtex4}
\usepackage{amssymb}
\usepackage{amsmath}
\usepackage{graphicx}
\usepackage{float}

\begin{document}

\title{Highly coherent spin states in carbon nanotubes coupled to cavity photons}

\author{T. Cubaynes${}^{1}$,  M.R. Delbecq${}^{1}$, M.C. Dartiailh${}^{1}$, R. Assouly${}^{1}$, M.M. Desjardins${}^{1}$, L.C. Contamin${}^{1}$, L.E. Bruhat${}^{1,2}$, Z. Leghtas${}^{1}$, F. Mallet${}^{1}$, A. Cottet${}^{{1}*}$ and T. Kontos${}^{1}$\footnote{To whom correspondence should be addressed: cottet@lpa.ens.fr, kontos@lpa.ens.fr}}
\affiliation{$^{1}$Laboratoire de Physique de l'Ecole normale sup\'{e}rieure, ENS, Universit\'{e} PSL, CNRS, Sorbonne
Universit\'{e}, Universit\'{e} Paris-Diderot, Sorbonne Paris Cit\'{e}, Paris, France}
\affiliation{$^{2}$Microtechnology and Nanoscience, Chalmers University of Technology, Kemiv\"{a}gen 9, SE-41296 Gothenburg, Sweden\\}

\date{\today}

\begin{abstract}
Spins confined in quantum dots are considered as a promising platform for quantum information processing. While many advanced quantum operations have been demonstrated, experimental as well as theoretical efforts are now focusing on the development of scalable spin quantum bit architectures. One particularly promising method relies on the coupling of spin quantum bits to microwave cavity photons. This would enable the coupling of distant spins via the exchange of virtual photons for two qubit gate applications, which still remains to be demonstrated with spin qubits. Here, we use a circuit QED spin-photon interface to drive a single electronic spin in a carbon nanotube based double quantum dot using cavity photons. The microwave spectroscopy allows us to identify an electrically controlled spin transition with a decoherence rate which can be tuned to be as low as $250kHz$. We show that this value is consistent with the expected hyperfine coupling in carbon nanotubes. These coherence properties, which can be attributed to the use of pristine carbon nanotubes stapled inside the cavity, should enable coherent spin-spin interaction via cavity photons and compare favourably to the ones recently demonstrated in Si-based circuit QED experiments.
\end{abstract}

\maketitle

\textbf{Introduction}

The observation of strong coupling between the charge or the spin confined in a quantum dot circuit and cavity photons has been reported very recently \cite{Bruhat:17,Mi:17,Stokklauser:17,Petta:18,Lieven:18,Wallraff:18}, bringing closer the demonstration of distant spin-spin interaction \cite{Trif:08,Cottet:10,Nori:12,Yacoby:18}, in the quest for scalable quantum information processing platforms \cite{LossDiVincenzo:98}. One critical parameter of a spin-photon platform is the linewidth of the spin transition which sets the maximum number of coherent swap operations between a spin and a photon. Whereas they are well documented in Si or GaAs, the coherence properties of single electronic spins in carbon nanotubes are still debated \cite{Laird:14,Laird:17}.

Here, we use a spin qubit scheme based on a carbon nanotube embedded in a microwave cavity. Our device is made using a stapling technique developed for cQED architectures, which produces ultraclean double quantum dot devices with near-ideal spectra. We use the circuit QED platform to perform a microwave spectroscopy of the spin transition. We observe the characteristic dispersion of the spin transition of our spin qubit. When the qubit states are tuned to be almost pure spin states (see figure 1d), the measured decoherence rate is found to be as low as $250kHz$. Strikingly, such a figure of merit is more than $100$ times better than in previous work on carbon nanotubes \cite{Laird:14,Laird:17} and compares favourably to the very recent values reported for Si based devices in a circuit QED environment\cite{Petta:18,Lieven:18}. From the gate dependence of the decoherence rate, we show that the charge noise is the main source of decoherence for the spin when the qubit states are mixed charge/spin states, but that it can be substantially reduced in the spin qubit regime.

The principle of our spin photon coupling relies on two non-collinear Zeeman fields on each quantum dots (see figure 1c and 1e) in a double quantum dot, originating from zig-zag shaped ferromagnetic contacts. These non-collinear Zeeman fields can be achieved by interface exchange fields \cite{Cottet:10} or by stray magnetic fields \cite{Nori:12,Benito:18} which both give the same hamiltonian. In our case, the interface exchange fields a priori dominate\cite{Viennot:15}. In the adiabatic regime, if an electronic spin is located on the left dot, it aligns along the left spin quantization axis whereas if it is located on the right dot, it aligns along the right spin quantization axis. Since the two dots are separated by few hundreds of nm, there is a large (mesoscopic) electric dipole between the left and the right dots which is given to the spin thanks to the non-collinear magnetisations. The photons of the cavity convey an electrical field which couples to this electric dipole and therefore to the spin. An alternative wording is to state that the ferromagnetic electrodes give rise to a two-site artificial spin orbit coupling, which makes the spin sensitive to the cavity electric field\cite{Viennot:15}.
It is interesting to note that such an ``orbitally'' mediated spin-photon coupling allows one to increase the natural spin-photon coupling by about 5 orders of magnitude \cite{Viennot:15,Petta:18,Lieven:18} without degrading substantially the inherent good coherence properties of a single spin if the device is used in the limit where the electron is trapped almost completely in one of the two dots (left or right). This regime can be reached by detuning the left dot orbital energy $\epsilon_L$ (tunable with gate voltage $V_L$) from the right dot  orbital energy $\epsilon_R$ (tunable with the gate voltage $V_R$). In such a gate configuration, the electron is trapped almost completely in one of the two dots , as sketched in figure 1e.
This allows to define the detuning $\epsilon_\delta=\alpha_L V_{L}-\alpha_R V_{R}+\epsilon_0$ along the white arrow of figure 2b, $\alpha_{L(R)}$ being given by the slope of the degeneracy line.
At the large detuning working point, the spin qubit is nearly insensitive to charge noise because of the almost flat energy dispersion but it keeps a coupling to the cavity field larger than the spin decoherence rate \cite{Cottet:10,Benito:18}.

\textbf{Results}

In the double quantum dot regime, each ferromagnetic contact in our device polarizes one quantum dot. This generates the spectrum depicted in figure 1d \cite{Cottet:10,Benito:18}. Each K/K' valleys of carbon nanotubes have a similar spectrum and we omit the valley index here. In addition, we study transitions outside the shaded grey region where the electronic states become pure spin states as the detuning is increased. Several features indicate very weak disorder and electrostatic control of the potential landscape of confined electrons via the bottom gates: a clear ``electron-electron quadrant'' delimited by the semiconducting gap controlled by two of the bottom gates ($V_{L}$ and $V_{R}$), continuous transition from double-dot spectrum (triple points and avoided crossings) to single dot spectrum (parallel transverse lines) and rather regularly spaced Coulomb blockade peaks. At the edges of the electron-electron quadrant of the stability diagram, we can form a double quantum dot in a controlled way in the few electron regime. We focus here on the zero detuning line between two charge states highlighted by a white square in figure 2a. The phase contrast of the microwave signal in this region is displayed on figure 2b. The phase of the microwave signal displays the characteristic sign change of a resonant interaction between the cavity and the double quantum dot. A transition to the dispersive (off-resonant) regime is also visible by a gradual change of phase contrast along this degeneracy line. In the resonant regime, the dependence of the phase contrast $\Delta \phi$ as a function of $\epsilon_\delta$ which has maxima/minima of about $\pm 15^{\circ}$ provides an estimate of the charge coupling strength $g_C \approx 2\pi (21 \pm 1 )MHz$ and of the charge decay rate $\gamma_C \approx  2\pi (1.35 \pm 0.16 ) GHz$ (see Supplementary Information).

The measured phase is determined by an average of the dispersive shifts induced by each transition, weighted by the steady-state occupation of each state (see fig 2b) (see e.g. \cite{Petersson:12,Viennot:14}). Applying a second tone allows to individually address the different transitions, and to recover their respective coupling strength to cavity photons.
The microwave spectroscopy of our ferromagnetic spin qubit is conveniently done by reading out in the dispersive regime the phase $\phi$ of the cavity signal when a second tone is applied through the cavity and its frequency is swept. At large detuning $\epsilon_\delta$, the phase is mainly sensitive to the expectation value  $\left\langle \sigma_z \right\rangle$ of the spin projection along the quantization axis of the left(right) dot. In the dispersive regime, the expression of the phase $\phi$ reads: $\phi = \frac{2g^{2}_{S}}{ \kappa \Delta } \left\langle\sigma_z \right\rangle +\phi_0$, where $\phi_0$ is a constant which only depends on the microwave setup, $\kappa$ is the linewidth of the cavity, $g_S$ is the spin-photon coupling strength and $\Delta=f_{cav}-f_{spin}$ is the detuning between the cavity frequency $f_{cav}$ and the spin qubit frequency $f_{spin}$. Such a measurement is shown in figure 3a which displays the phase contrast $\Delta \phi$ as a function of the tone frequency $f_{pump}$ and $\epsilon_\delta$. In order to avoid cavity photon back-action on the spin qubit, we use a pulsed microwave spectroscopy with the pulse sequence shown in figure 3c. The qubit is first driven for $t=3 \mu s$, then the cavity is filled after $90ns$ and finally read-out using a fast data acquisition card for $t=700ns$. Apart from the frequency independent vertical blue stripe which simply signals the left/right degeneracy line at zero detuning, we observe 3 resonances which disperse close to zero detuning and saturate at $6.506 GHz$, $6.530 GHz$ and $6.540 GHz$ respectively. The dispersion of each of these transitions with a minimum at zero detuning and a saturation at large detuning is characteristic of a transition which becomes a pure spin transition in the large detuning limit due to the perfect localization of the electron in one quantum dot (see figure 1d). The saturation value is given by the effective Zeeman field felt by the (pure) spin state at large detuning. The fact that we observe several spin transitions can be attributed to the lifting of the K/K' valley degeneracy of the nanotube as well as from the fact that we are not in the single electron regime. As expected for a spin transition, we can tune the value of this saturation with the external magnetic field. The resulting dependence is shown in figure 3b. The low slope is consistent with previous measurements in a similar architecture with non-stapled nanotube material \cite{Viennot:15} and could arise from field modulated exchange coupling between the dot's spin and its adjacent ferromagnetic contact \cite{Cottet:11} (see Supplementary Information).

A cut along the lowest resonance at large detuning is shown in figure 3d. This measurement fitted by a lorentzian has a full width at half maximum of $\gamma_{FWHM} = 2\pi \times (498 \pm 80)kHz$ which sets an upper bound of the decoherence rate $\gamma_S \leq \gamma_{FWHM} /2 = 2\pi \times 249 kHz$. Such a narrow line width is two orders of magnitude lower than that found in the valley-spin qubit in previous work with carbon nanotubes\cite{Pei:12} and compares favorably to the very recent figures of merit reported for Si based platforms \cite{Petta:18,Lieven:18}. We speculate that such a figure of merit is due mainly to the purity of our nanotube-metal interfaces. From the phase contrast of about $4^{\circ}$, we can estimate a lower bound spin photon coupling strength $g_{S}\approx 2\pi \times (2.0 \pm 0.1) MHz$ (see supplementary), which exceeds the decoherence rate of the spin states (and of the cavity) and therefore implies that the spin is strongly coupled to the cavity photons although they are not resonant. As a comparison, we have $g_S /\gamma_S \geq 8$ for the spin transition whereas $g_C/\gamma_C\approx 0.015$ for the charge like transition (see supplementary). There is therefore a very large gain in the coherence properties of our device when we switch from the charge like transitions to the spin transitions.

\textbf{Discussion}

In order to specify the decoherence mechanism explaining the linewidth found for our spin transition, we have measured the dependence of the decoherence rate as a function of the detuning $\epsilon_\delta$. Such a measurement is displayed in figure 4a. Two main decoherence sources are expected for the electronic spin in double quantum dots: charge noise and nuclear spin. Our ${}^{12} C$ platform is grown from a natural $CH_4$ feedstock gas and is thus expected to have a low concentration of nuclear spins ($1.1\%$ of ${}^{13} C$). The charge noise is related to the fact that the qubit transition frequency may fluctuate if offset charges nearby the device change the detuning. Therefore, it should induce a decoherence rate $\gamma_S$ proportional to the derivative of the qubit transition frequency with respect to the detuning\cite{Cottet:10}. For a large detuning $\epsilon_\delta$, the nuclear spin bath is on the contrary expected to give a nearly independent contribution as a function of the detuning. The decoherence rate $\gamma_S$ and the derivative $\partial \omega /\partial \epsilon_\delta$ as a function of detuning $\epsilon_\delta$ are shown to overlap well provided we add a residual constant of about $500 kHz$ to the derivative in figure 4a. The linear behavior of the decoherence rate $\gamma_S$ as a function of the derivative $\partial \omega /\partial \epsilon_\delta$, displayed in inset, shows that our spin-photon interface is dominated by charge noise at small detuning. Interestingly, it allows us from the slope of the linear behavior to extract a charge noise detuning variance of about $34 \mu eV$. While this noise is larger than in previous work in carbon nanotubes \cite{Viennot:14} and could be in principle easily lowered, it is interesting to see that we can completely reduce its influence by going at large detuning while keeping a large spin-photon coupling strength with respect to $\gamma_S$. The shaded grey corresponds to the residual decoherence mechanisms with a decoherence rate in the range $\approx 2\pi \times 500kHz$. Note that this value corresponds to twice as much as the lowest decoherence rate presented in figure 3d ($250kHz$), probably because it corresponds to a lower detuning. Interestingly, the residual decoherence rate allows us to give an upper bound of the contribution of the nuclear spins of the $1.1 \%$ of ${}^{13} C$ and therefore of the hyperfine coupling $\mathcal{A}$. From the estimated diameter of our CVD nanotubes $D\approx 2 nm$ and the length of each dot $d \approx 500nm$, we get a number of nuclear spins of $N \approx 0.011 \times 3\times 10^5$ which yields $\gamma_S\approx 2\pi \times 200 kHz$ if $\mathcal{A}=0.1\mu eV$. Our measurements are therefore fully in agreement with the tabulated values for the hyperfine coupling expected in CNTs of $\mathcal{A}\approx 0.1-0.5 \mu eV$ (see ref. \cite{Laird:17} and reference therein). In addition to the decoherence rate, we also present the spin-photon coupling strength and the cooperativity of the spin-photon interface ($C=(2g_{S}^2)/(\kappa \times \gamma_S)$) as a function of the detuning (figure 4b and 4c respectively). Interestingly this last quantity allows to identify an optimal detuning working point around $\epsilon_\delta=-18 GHz$. For this detuning, the hybridization with the charge creates a sizeable spin-photon coupling while maintaining a low decoherence rate \cite{Lieven:18}. Our results suggests that carbon, like silicon, can be a promising host for electronic spins encoding quantum information. This is enabled by our clean and controlled nano-assembly technique of carbon nanotubes in cavity and could be further enhanced by purified ${}^{12} C$ growth to get rid of the nuclear spins.

In summary, we have demonstrated that carbon nanotube based double quantum dots can provide a tunable and coherent spin-photon interface. The figures of merit of coupling strength of $g_{S}\approx 2\pi \times 2.0 MHz$ and low decoherence rate $\gamma_{s}\approx 2\pi \times 250kHz$ are suitable for future swap experiments. It could be interesting in that context to increase the value of $g_S$. We anticipate that this could be done by optimizing the angle between the ferromagnets\cite{Cottet:10,Benito:18} or by increasing the impedance of our $50 \Omega$ resonator similarly to Refs.\cite{Petta:18,Lieven:18}.

\textbf{METHODS}\\
Our devices are made with a complete dry transfer nanotube technique adapted from previous works \cite{Zhong:11,Pei:12,Waissman:13,Viennot:14b,Ranjan:15} which allows us to integrate as-grown carbon nanotubes in a microwave cavity. The full chip, comprising the cavity, the bottom gates and the non-collinear ferromagnetic contacts shown in figure 1a, 1b and 1c respectively, is placed in a vacuum chamber with a base pressure of $5\times 10^{-7} mbar$. The zig-zag contacts visible in figure 1b and partially in figure 1c are NiPd ferromagnetic contacts with transverse magnetization\cite{Viennot:15}. Carbon nanotubes are grown on a Si comb with a standard Chemical Vapor Deposition (CVD) recipe with $CH_4$ as feedstock gas. The comb is mounted inside the chamber on a stage with micro- and nano-manipulators which allow us to place the nanotube on the chip with controlled approach steps of about $100nm$. The assembly of the carbon nanotube and the ferromagnetic contacts is done under vacuum at a pressure of about $1\times 10^{-6} mbar$ in order to ensure a clean interface between the nanotube and the metallic contacts. This results in the device shown in figure 1c where a nanotube bridges the two ferromagnetic contacts and is \emph{a priori} suspended over bottom gates. The wider gate visible in the SEM picture of figure 1c is galvanically coupled to the central conductor of the Nb cavity visible in figure 1a. The cavity fundamental resonance frequency is $6.424 GHz$ and its quality factor is about $4200$. 

The devices obtained with our fabrication technique are more tunable than previous nanotube based spin quantum bits and much less disordered \cite{Viennot:15}. The measurement setup is similar to the setup of ref. \cite{Viennot:15}. We measure simultaneously the DC current $I$ flowing through the device and the microwave signal transmitted through the cavity in amplitude $A$ and phase $\phi$. The control which we have on the spectrum of the device is visible from the stability diagram shown in figure 2a which displays the current under a bias of $V_{sd}=100 \mu V$. The horizontal axis of the microwave spectroscopy is calibrated using the triangles observed in the transport data (not shown). The calibrated parameters are the charging energies $E_{cR}=2.36meV$, $E_{cL}=2.21meV$, $U_{m}=0.61meV$ and the gate capacitances $C_{gL}=1.6aF$, $C_{gR}=0.9aF$.

{\noindent\small{\textbf{Data availability} }Supplementary Information is available at npj Quantum Information website. The datasets generated during and/or analysed during the current study are available from the corresponding authors on reasonable request.

{\noindent\small{\textbf{Acknowledgements.} } The devices have been made within the consortium Salle Blanche Paris Centre. We gratefully acknowledge help from Jos\'e Palomo, Aur\'elie Pierret and Michael Rosticher. This work is supported by the ERC Starting Grant "CirQys" and by the ANR "FunTheme".

{\noindent\small{\textbf{Competing financial interests.} } The authors declare no competing financial interests.

\begin{figure}[!ht]
\centering\includegraphics[width=0.8\linewidth,angle=0]{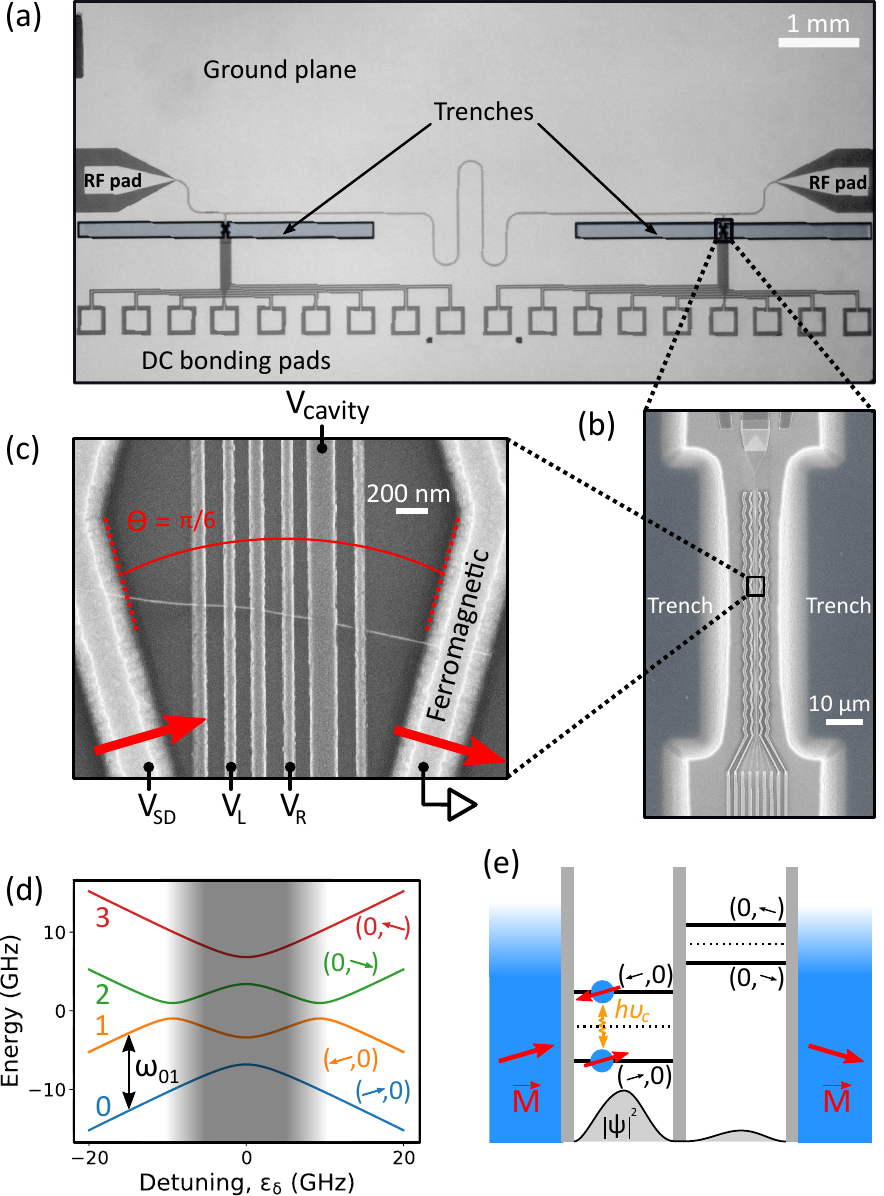}
\caption{\textbf{Principle of spin-photon interface} \textbf{a.} Large scale view of our circuit QED setup. Only one of two circuit areas is used. \textbf{b.} Scanning electron microscope (SEM) picture of the pedestal structure on which the device is made with the zig-zag magnetic contacts. \textbf{c.} Zoom on the device showing the bottom gates and the non-collinear ferromagnetic contacts. \textbf{d.} Spectrum of our spin quantum bit. The spin transition addressed in this work, the 01 transition, saturates to a value defined by the effective Zeeman splitting of each dot at large detuning. In the shaded grey region, spin and charge are not good quantum numbers anymore. \textbf{e.} Schematics of our device showing the concept of spin-photon coupling.
}%
\label{device}
\end{figure}

\begin{figure}[!ht]
\centering\includegraphics[width=0.8\linewidth,angle=0]{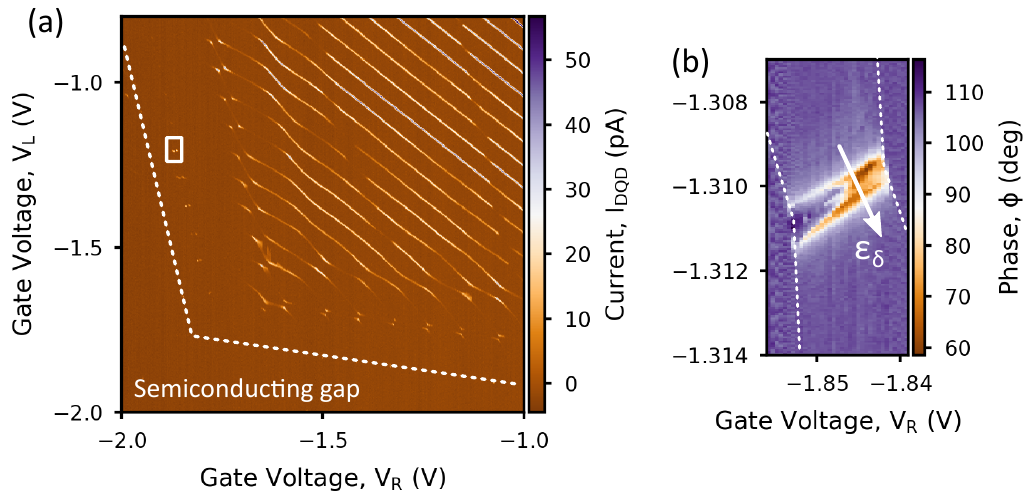}
\caption{\textbf{Carbon nanotube double quantum dot in the few electron regime} \textbf{a.} Stability diagram of our double quantum dot spin qubit, given by the current I through the double dot versus the gate voltages $V_{L}$ and $V_{R}$. The gap is at the lower left corner of the stability diagram and is indicated by white dashed lines. \textbf{b.} Phase contrast for the transmitted microwave signal along a charge degeneracy line in the region indicated by the white square in figure 2a.
}%
\label{greyscale}%
\end{figure}

\begin{figure}[!ht]
\centering\includegraphics[width=0.8\linewidth,angle=0]{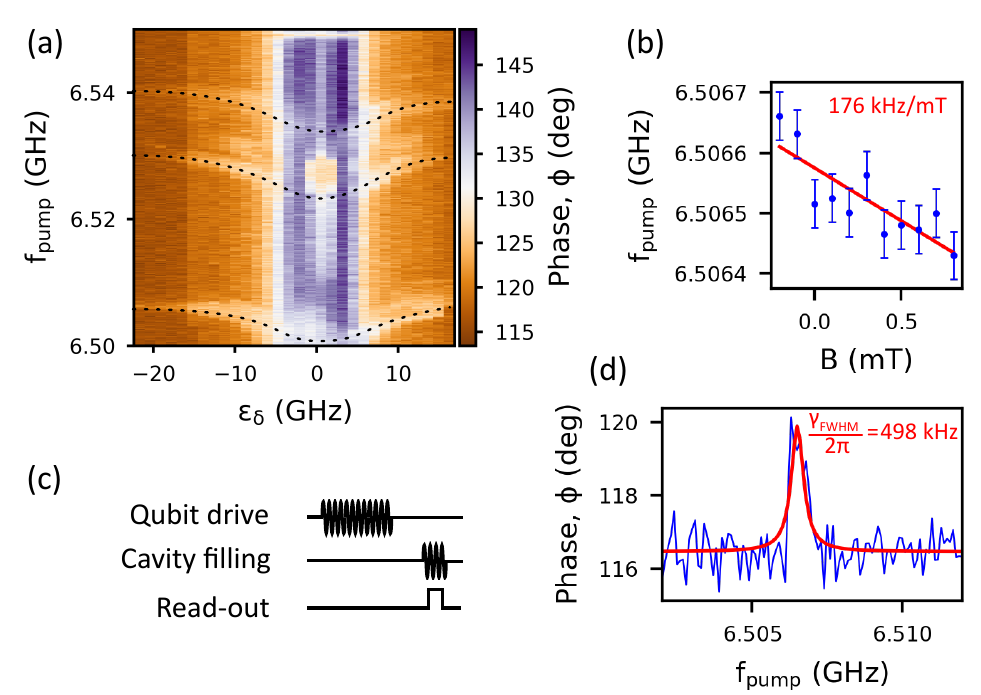}
\caption{\textbf{Microwave spectroscopy of spin transition} \textbf{a.} Microwave two-tone spectroscopy. \textbf{b.} Magnetic field dependence of the microwave frequency (because of setup constraints, the magnetic field span was limited to $\pm 1 mT$ in this experiment). \textbf{c.} Pulse sequence used for the two-tone spectroscopy. \textbf{d.} Resonance corresponding to the lowest transition of figure 3a at a large detuning $<-21GHz$ (blue line). We cannot give the exact detuning value due to a gate offset jump which occurred during the measurement. The red line is the lorentzian fit which allows us to extract $\gamma_{FWHM}$.
}%
\label{coupling}%
\end{figure}

\begin{figure}[!pth]
\centering\includegraphics[width=0.8\linewidth,angle=0]{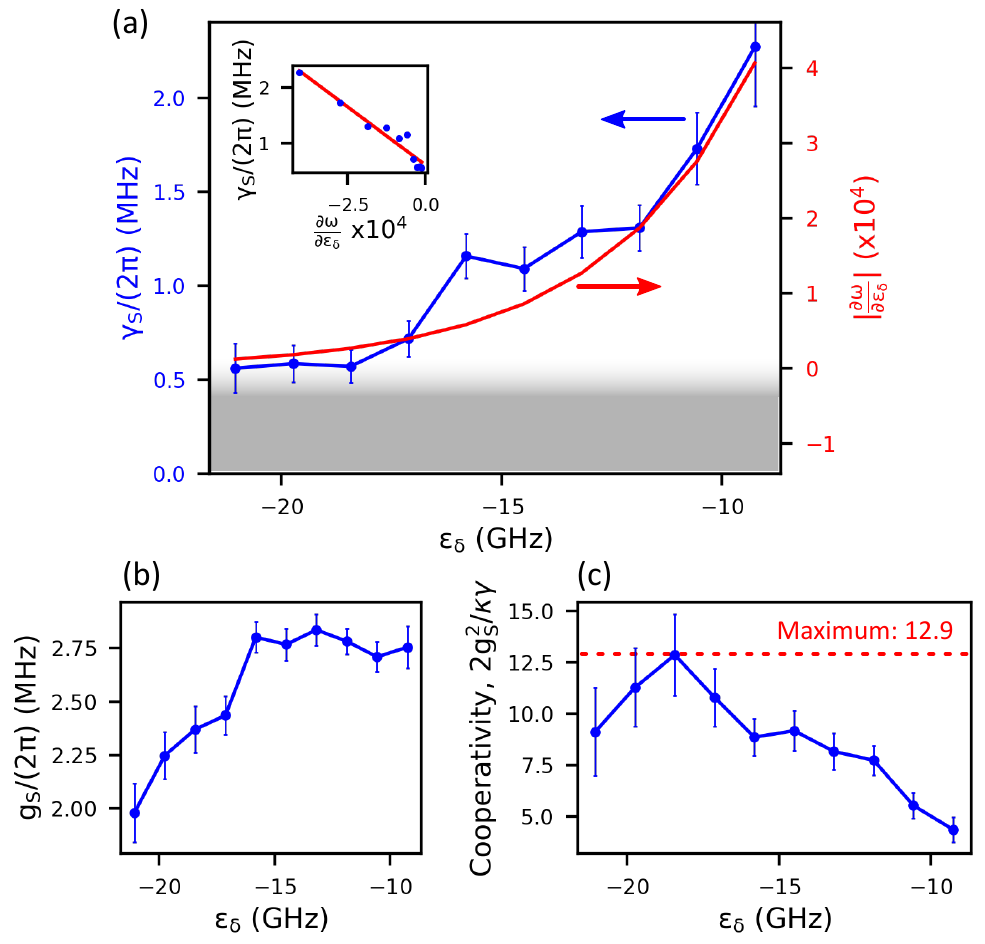}
\caption{\textbf{Nuclear spin limited spin photon interface} \textbf{a.} Linewidth and derivative of the dispersion relation of the spin transition as a function of detuning (the derivative is obtained by fitting the dispersion relation of Fig.3a and calculating the derivative from the fit). A constant corresponding to a decoherence rate of 560kHz is added to the derivative. Inset : Linewidth as a function of derivative. \textbf{b.} Spin-photon coupling strength as a function of detuning. \textbf{c.} Spin-photon cooperativity as a function of detuning.
}%
\label{coupling}%
\end{figure}

\end{document}